\documentclass[final,twocolumn]{elsarticle}
\usepackage{amssymb}
\UseRawInputEncoding

\begin{document}

\begin{frontmatter}

\title{First-order transition in the stacked-$J_1$-$J_2$ Ising model on a cubic lattice}

\author{A.O. Sorokin}
\ead{aosorokin@gmail.com}

\address{Petersburg Nuclear Physics Institute, NRC Kurchatov Institute, 188300 Orlova Roscha, Gatchina, Russia}

\begin{abstract}
We investigate critical properties of the stacked-$J_1$-$J_2$ Ising model on a cubic lattice. Using Monte Carlo simulations and renormalization group, we find a single phase transition of the first order for $J_2/J_1>1/2$. The renormgroup approach predicts that a transition can be of the second order from the universality class of the $O(2)$ model, but the Monte Carlo results show another set of critical exponents: exponents continuously vary form the values typical for a first-order transition in the finite-size scaling theory at $1/2<J_2/J_1<1$ to the Ising values in the limit $J_2/J_1\to\infty$. We also exclude the pseudo-first-order behavior observed in the $J_1$-$J_2$ Ising model on a square lattice for $0.67\lesssim J_2/J_1\lesssim0.9$.
\end{abstract}

\end{frontmatter}

First-order (discontinuous) and continuous phase transitions differ in the presence or absence of a jump in the order parameter and internal energy, as it is reflected in the title of a transition type. Despite the clarity of this definition, in practice, a transition type can not always be determined clearly and unambiguously, even using the most reliable theoretical methods.  At that, both situations are realized: when a first-order transition seems like a continuous one, and vise versa.

In the first situation, it is customary to say about the weak first order. In terms of the renormalization group (RG), such a situation arises if a RG-trajectory passes through a vicinity of a saddle fixed point or a fixed point with complex-valued coordinates with a small imaginary part \cite{Zumbach93}. Herewith, a RG-flow is rather slow, so a transition does not show first-order sings in a wide range of scale (or lattice size), and furthermore pseudo-scaling behavior may be observed. Sings of the first order may appear only when one consider lattices with large sizes. Moreover, the weakness of the first order may lead to that various theoretical approaches not using finite-size lattices predict different types of a transition. Perhaps, the most vivid example of this is a Heisenberg antiferromagnet on a stacked-triangular lattice, where the $4-\varepsilon$ expansion \cite{Sokolov20} and the non-perturbative RG \cite{Delamotte16} predict a first-order transition, while the perturbative RG \cite{Pelissetto01,Sokolov02} and the conformal bootstrap \cite{Nakayama14,Henriksson20} show the second order. In this case, even various Monte Carlo methods predict different results, so the results \cite{Diep08} based on the Wang-Landau algorithm \cite{Wang01} confirm the first order, while the recent results \cite{Kawamura19} based on the combining of the Metropolis and over-relaxed algorithms show the second-order behavior (but the same method predicts a first-order transition for frustrated helimagnet \cite{Sorokin14}).

The second situation corresponds to the so-called pseudo-first order, when the transition exhibits first-order behavior on small lattices, but the first-order signs disappear in the thermodynamic limit. In terms of RG, such a situation arises if a RG-trajectory starting in the region of a stable potential passes through the stability region boundary but tends to the fixed point locating inside the region or on the boundary \cite{Sorokin21}. A pseudo-first-order transition has been previously found in two-dimensional models: 4-state Potts model and the Ashkin -- Teller model \cite{Salas96}, the Wu -- Baxter model \cite{Adler05} and the $J_1$-$J_2$ Ising model on a square lattice \cite{Jin12,Kalz12,Jin13}. (Such a behavior is also discussed for five- and higher-dimensional models \cite{Lundow11,Sorokin21}.) The later model is interesting because its three-dimensional analogue, namely the stacked-$J_1$-$J_2$ Ising model, has a transition a type of which is still controversial.

\begin{figure}[t]
\center
\includegraphics[scale=0.30]{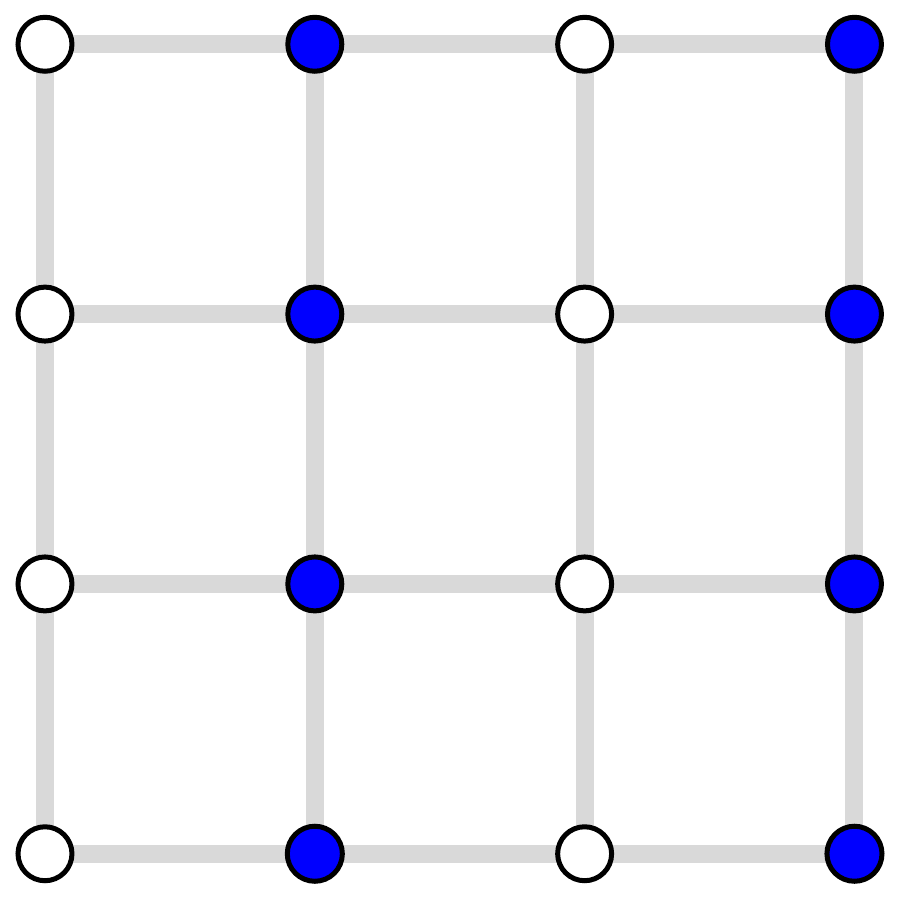}\hspace{9mm}
\includegraphics[scale=0.30]{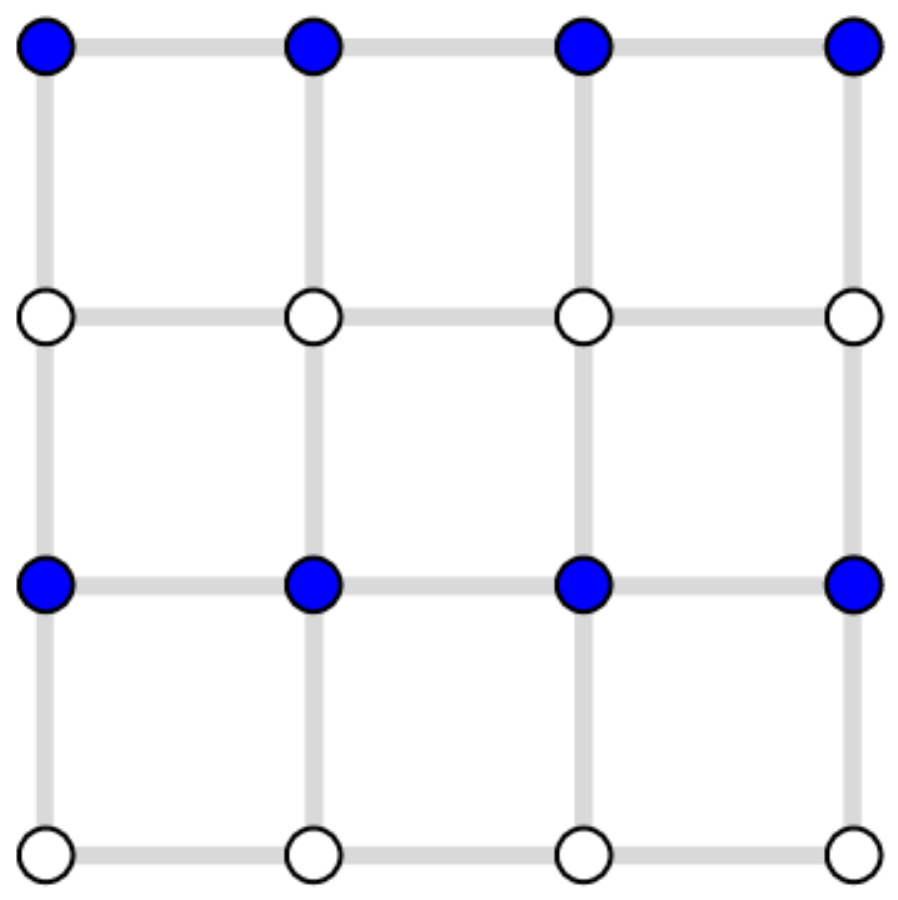}\vspace{9mm}

\includegraphics[scale=0.30]{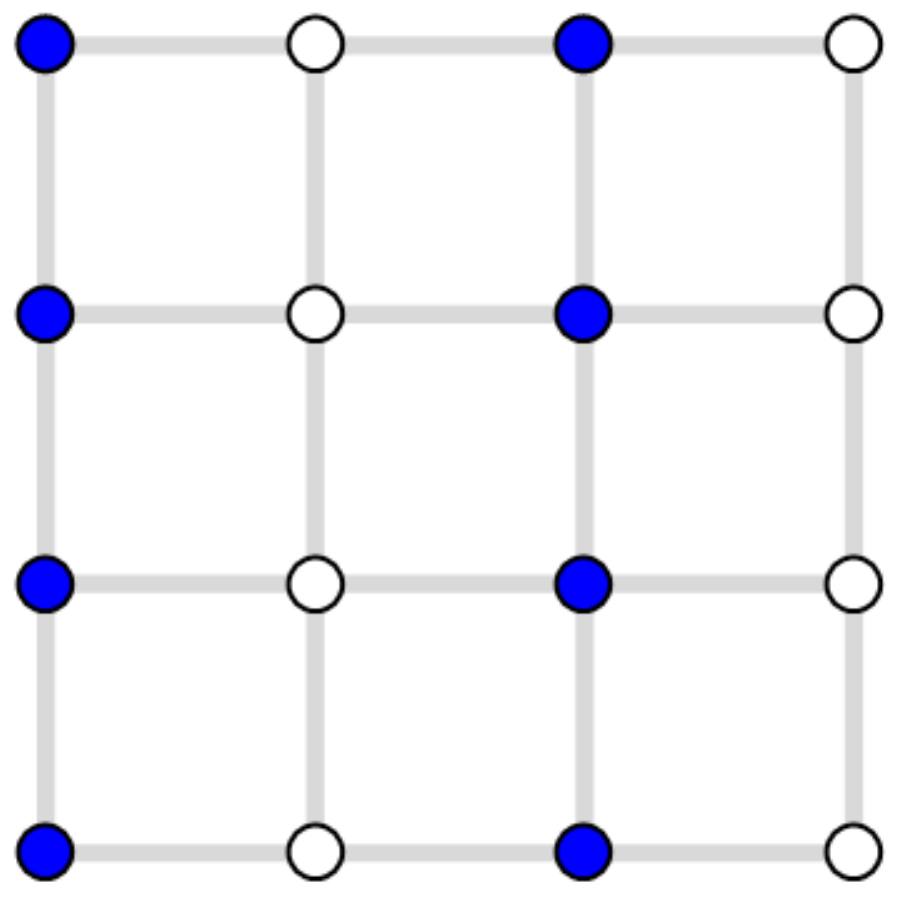}\hspace{9mm}
\includegraphics[scale=0.30]{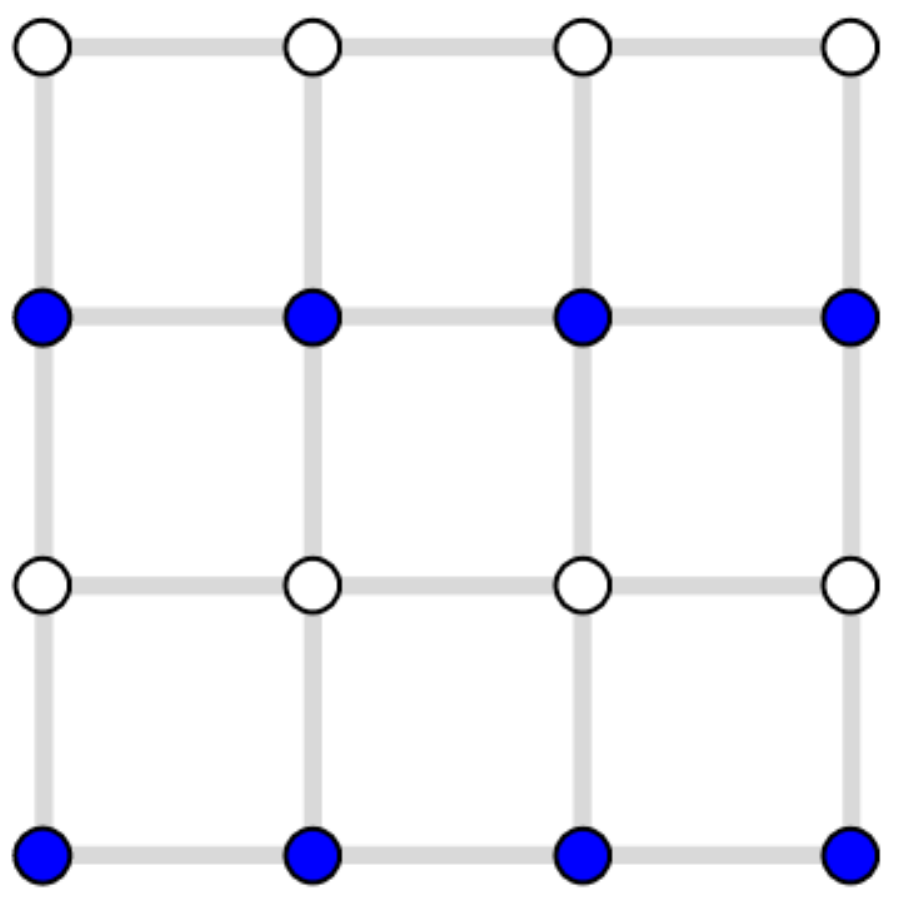}
\caption{\label{fig1}Four ground state configurations.}
\end{figure}%
The $J_1$-$J_2$ Ising model is an Ising antiferromagnet with the additional interaction between next-nearest-neighbor spins on a square lattice. The model and its three-dimensional generalization are described by the Hamiltonian
\begin{equation}
    H=J_1\sum_{ij}s_i\cdot s_j+J_2\sum_{kl}s_k\cdot s_l+J\sum_{mn} s_m\cdot s_n,
    \label{model}
\end{equation}
where $s=\pm1$, the sum $ij$ runs over pairs of nearest-neighbor spins in layers, the sum $kl$ enumerates pairs of next-nearest-neighbor spins in layers, and the sum $mn$ runs over pairs of nearest-neighbor spins in the interlayer direction. The signs of the exchange constants $J_1$ and $J$ do not affect the critical behavior, so for simplicity we fix $J_1=J>0$. The frustrated state appears only if $J_2>0$. At $J_2<J_1/2$, the ground state is the non-frustrated antiferromagnetic order. At $J_2>J_1/2$, the ground state is one of two spin configurations with the wave-vectors $\mathbf{q}=(\pi,0,\pi)$ or $\mathbf{q}=(0,\pi,\pi)$. For Ising spins, the four possible ground state configurations are shown in fig. \ref{fig1}.
This model can be interpreted as two Ising models on $\sqrt2\times\sqrt2$ sublattices interacting with the coupling constant $J_1$, so the limit $J_2/J_1\to\infty$ corresponds to two non-interacting Ising models.

Another model equivalent to two interacting Ising models is the Ashkin-Teller model \cite{Fan72}. In two dimensions ($J=0$), the phase diagram of this model contains the (self-dual) line of transitions with varying critical exponents \cite{Kadanoff79,Kadanoff79-2,Kadanoff80}. The possibility of continuous variation of exponents is realized for critical points whose conformal symmetry has the central charge $C\geq1$ \cite{Zamolodchikov86,Zamolodchikov87}. The case $C=1$ corresponds, e.g., to the Ashkin -- Teller, 4-state Potts, $O(2)$ (free boson) and two Ising models. This transition line has the end point corresponding to the 4-state Potts model and the intermediate point corresponding to decoupled Ising models. The section between these two special points is mapped to the section of the $J_1$-$J_2$ model for $J_2/J_1\geq g^*$, where $g^*\approx0.67$ \cite{Jin12,Kalz12,Jin13}. For $1/2\leq J_2/J_1<g^*$, a transition is of the first order. In previous works, the position of the tricritical point has been estimated as $g^*\approx0.9$. To exclude the pseudo-first-order behavior using Monte Carlo simulations, the authors  \cite{Jin12,Kalz12,Jin13} have been compelled to consider large lattices (the signs of a first-order transition have been observed for lattice sizes $128\leq L<2000$). Besides Monte Carlo simulations, in the work \cite{Jin13} it has been performed the calculation using the cluster mean-field approach which is confirmed the position of the tricritical point $g^*\approx0.66$.

Perhaps, it is useful to note that the pseudo-first-order behavior in two-dimensional lattice models can be disclosed without a consideration large-size lattices by an estimation of the central charge value. The geometrical meaning of the conformal central charge is the Casimir energy of a cylindrycal form of a lattice, where effects of the pseudo-first-order behavior are subdued with increasing the ratio of the length and radius of a cylinder. The rather effective method for this applicable to the $J_1$-$J_2$ model has been proposed in \cite{Sorokin17}.

As other mean-field methods, the cluster mean-field approach does not take into account the influence of critical fluctuations significant at least in dimensions less than four (for higher dimensions see, however \cite{Sorokin21} and refs. therein), where these fluctuation may induce a first-order transition even if the Landau or other mean-field theories predict a continuous one. So, the position of the tricritical point in the $J_1$-$J_2$ model is caused not by critical fluctuation but other reasons, perhaps connected with the two-dimensional conformal symmetry.

\begin{figure}[t]
   \center
   \includegraphics[scale=0.28]{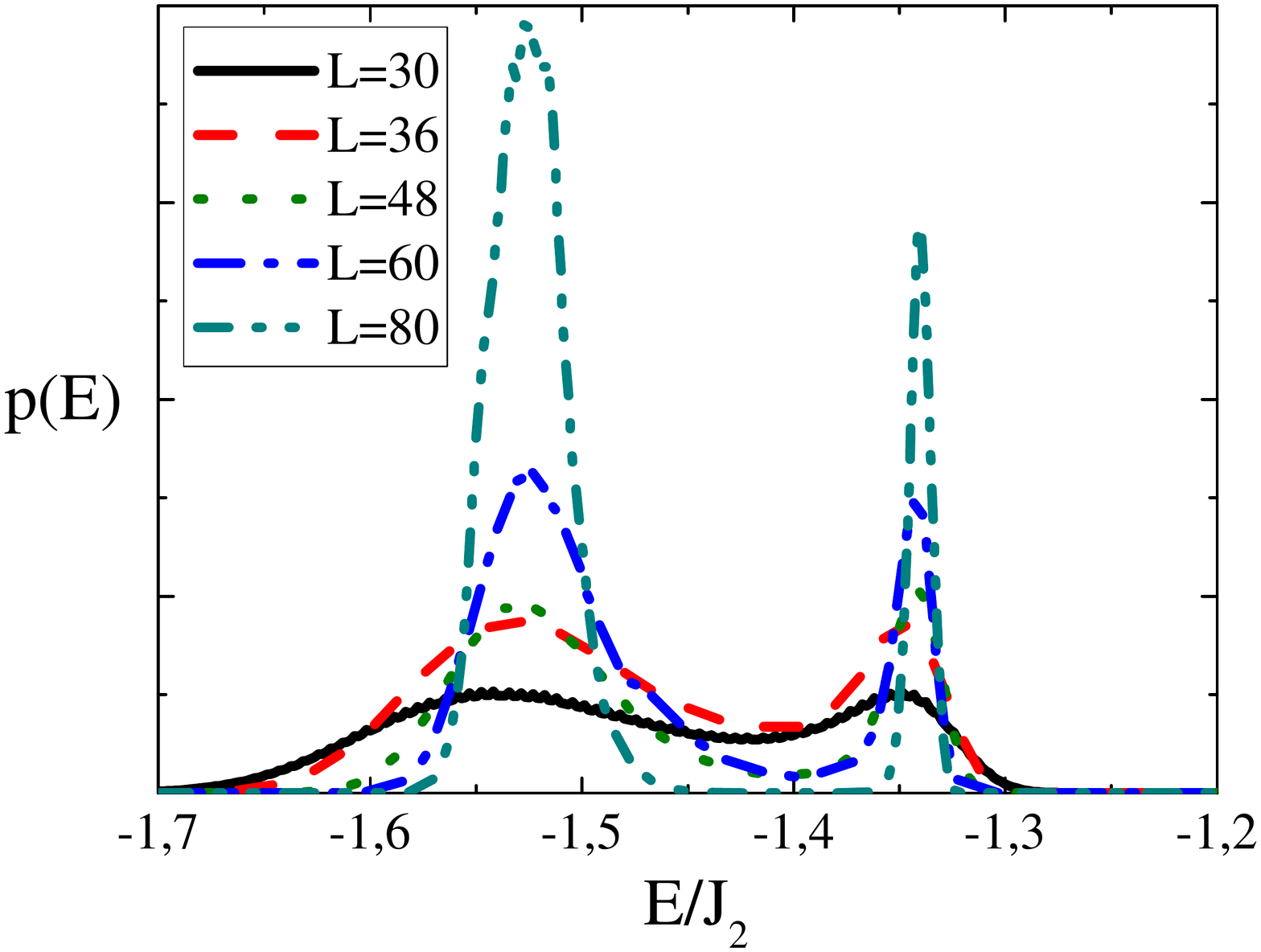}
   \caption{\label{fig2}Energy distribution near the transition temperature at $J_2/J_1=2/3$. For $L=30$, $T/J_2=3.6950$; $L=36$, $T/J_2=3.6950$; $L=48$, $T/J_2=3.6957$; $L=60$, $T/J_2=3.6957$; $L=80$, $T/J_2=3.6960$.}
\end{figure}%
In the recent work \cite{Godoy20}, the cluster mean-field approach has been used for the generalization to the three-dimensional case $J\neq0$. The authors find that the tricritical point is absent in the frustrated case $J_2/J_1>1/2$, so a transition is of the second order. This result contradicts to the Monte Carlo simulations performed using the replica exchange algorithm \cite{Murtazaev15,Murtazaev16,Murtazaev17}, where a first-order transition is found for $1/2<J_2/J_1<0.9$, and to the results obtained with the Wang -- Landau algorithm, where a first-order transition is found even for $J_2/J_1=1$ \cite{Sorokin18}. The authors \cite{Godoy20} argue that their method successfully predicts the position of the tricritical point in two dimensions $J=0$, and a first-order transition found by the Monte Carlo simulations can turn out to be of the pseudo-first-order, since lattice sizes $L\leq90$ considered in the simulations  are not sufficiently large to detect it.

To resolve this contradiction, one should use methods which correctly take into account critical fluctuations. For the numerical part of this work, we choose the Metropolis algorithm, though it is not the most efficient but it is universal and gives easily reproducible results. As an analytical method, we use RG with the $4-\varepsilon$ expansion. Below, we give the arguments that a transition in the frustrated case $J_2/J_1>1/2$ of the stacked-$J_1$-$J_2$ model (at least with $J/J_1=1$) is of the first order.
        \begin{figure}[t]
            \center
            \includegraphics[scale=0.28]{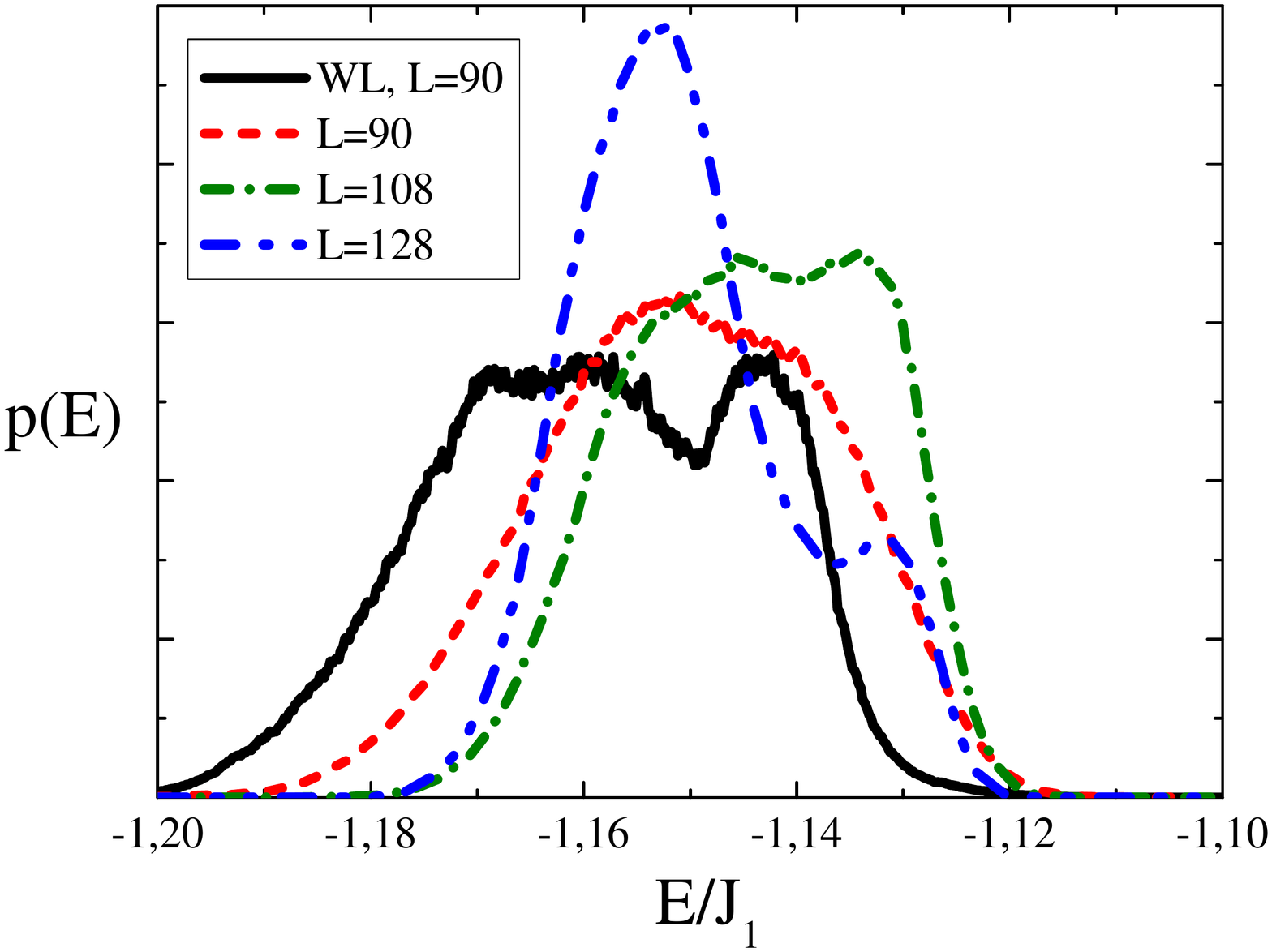}
            \caption{\label{fig3}Energy distribution near the transition temperature at $J_2/J_1=1$. For $L=90$ and the Wang -- Landau algorithm $T/J_1=4.1759$ \cite{Sorokin18}; for the Metropolis algorithm and $L=90$, $T/J_1=4.175$; $L=108$, $T/J_1=4.176$; $L=128$, $T/J_1=4.1767$.}
        \end{figure}%

\begin{enumerate}
    \item
        At $J_2/J_1=2/3$, a transition is of the distinct first order, and the pseudo-first-order behavior does not observed. For the later case, the difference between the peaks position in the energy distribution $\Delta E_\mathrm{peaks}$ as a function of a lattice size $L$ is expected to be
        \begin{equation}
            \Delta E_\mathrm{peaks}(L)\sim L^{-d/2}.
        \end{equation}
        The fig. \ref{fig2} shows that the peaks position does not change for a rather wide range of a lattice size $30\leq L\leq80$.

    \item
        At $J_2/J_1=1$, a transition is of a weak first order. The Metropolis algorithm for a rather large lattices $90\leq L\leq128$ also detect the double-peak energy distribution, though it is almost elusive for these lattice sizes (see fig. \ref{fig3}). Note that the previous result \cite{Sorokin18} using the Wang -- Landau algorithm is consistent with this result but still has a small mismatch. Such a inconsistency appears due to our inaccuracy in the realization of the algorithm (a choice of the histogram flatness condition, a number of iteration an so on), but, of cause, it is not a defect of the algorithm.

    \item The Fisher critical exponent is negative for all $J_2/J_1>1/2$, $\eta<0$ (see fig. \ref{fig4}) that indicates a first-order transition \cite{Pokrovskii}. To estimate critical exponents (or pseudo-exponents for a weak first-order transition), we perform Monte Carlo simulations for $J_2/J_1=2/3,\,1,\,4/3,\,2,\,5,\,10$. We use periodic boundary conditions and consider lattice sizes $16\leq L\leq 48$. Thermalization to an equilibrium state is performed within $6\cdot10^5$  Monte Carlo steps per spin, and calculation of averages within $9\cdot10^6$ steps. One finds details of the simulation technics in \cite{Sorokin19-2}. Note that our results are in agreement with the results \cite{Murtazaev15,Murtazaev16,Murtazaev17}. In particular, the results for the case $J_2/J_1=1$ are shown in table \ref{tab1}.
        \begin{figure}[t]
            \center
            \includegraphics[scale=0.28]{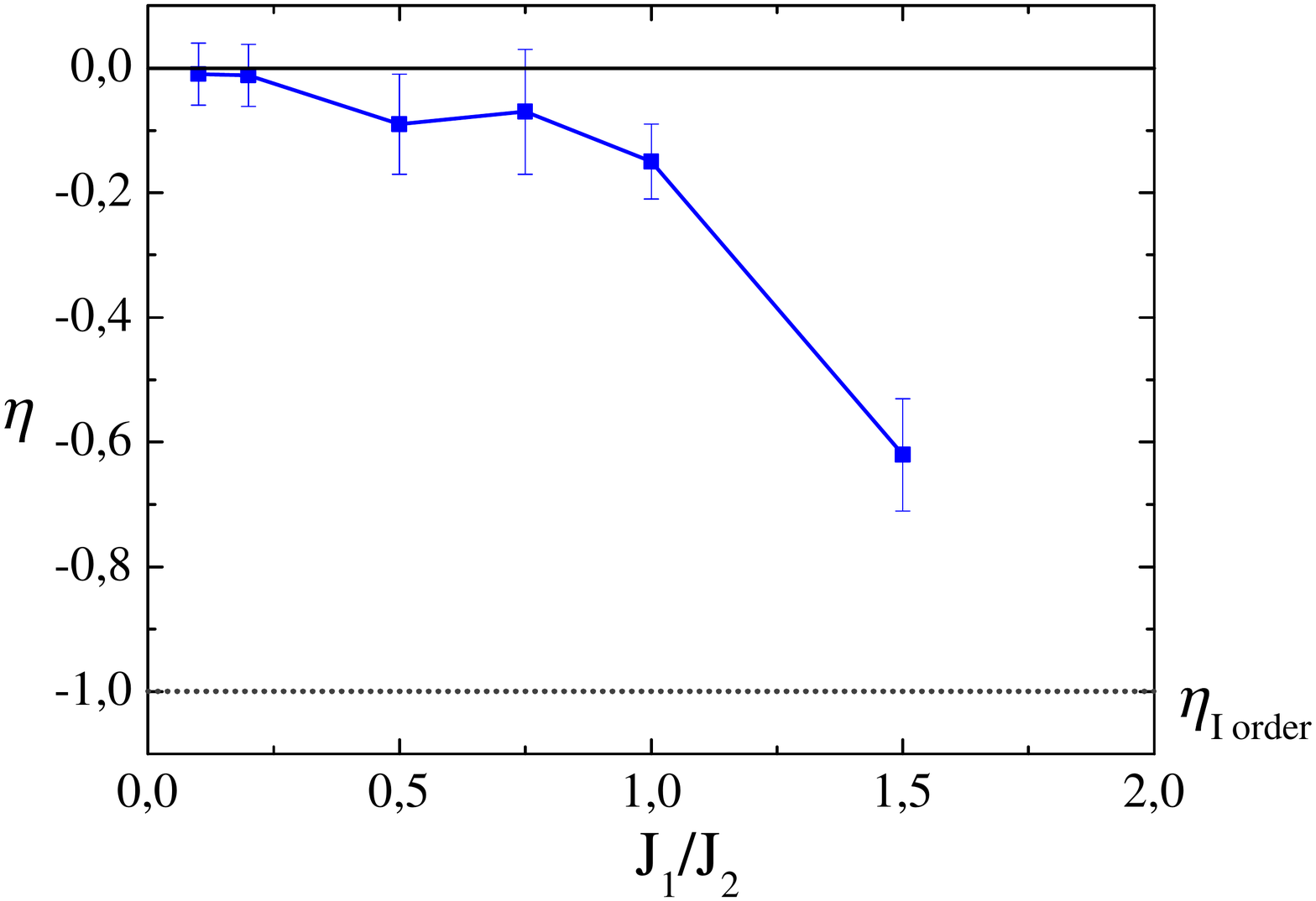}
            \caption{\label{fig4}Dependence of the critical exponent $\eta$ on the coupling constant ratio $J_1/J_2$.}
        \end{figure}%
        \begin{table}[b]
            \center
            \caption{\label{tab1}Critical (pseudo)exponents at $J_2/J_1=1$. The results of this work mark as [*].}
            \begin{tabular}{ccccc}
                \hline
                \hline
                 & $\nu$ & $\beta$ & $\gamma$ & $\eta$\\
                \hline
                [*]          & 0.536(6) & 0.230(6) & 1.15(2) & -0.15(6)\\
                \cite{Murtazaev15} & 0.549(5) & 0.245(5)& 1.190(5) & -0.16(2)\\
                \hline
                \hline
            \end{tabular}
        \end{table}

        \begin{figure}[t]
            \center
            \includegraphics[scale=0.28]{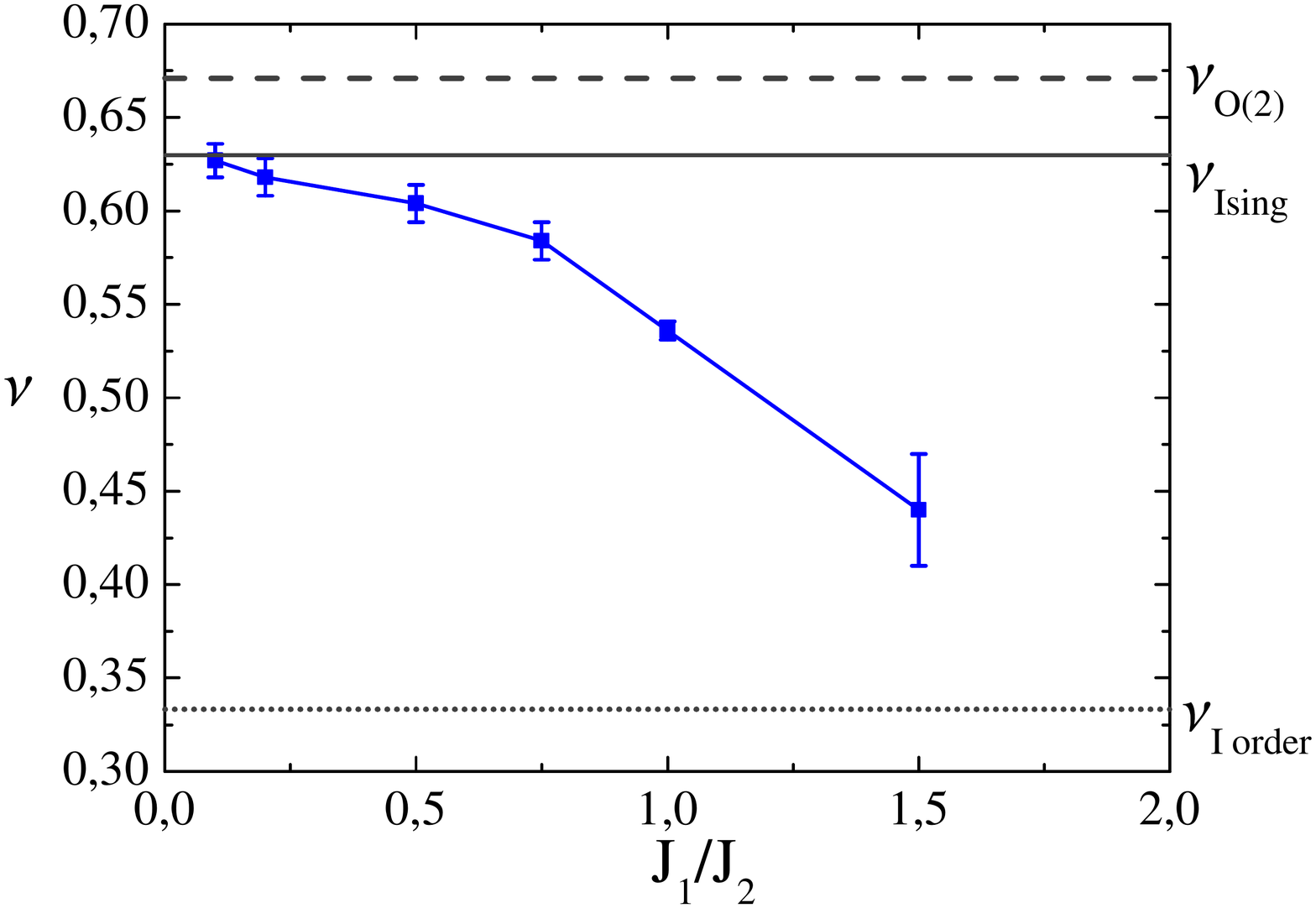}
            \caption{\label{fig5}Dependence of the critical exponent $\nu$ on the coupling constant ratio $J_1/J_2$.}
        \end{figure}%
    \item The $4-\varepsilon$ expansion predicts either a transition is of the second order from the universality class of the $O(2)$ model or a fluctuation induced first-order transition, but critical (pseudo-)exponets estimated by Monte Carlo simulations do not correspond to this universality class. Fig. \ref{fig5} (see also table \ref{tab1}) shows that the exponent $\nu$ does not have the value $\nu\approx0.67$ typical for the universality class of the $O(2)$ model for all $J_2/J_1>1/2$, but tends to the Ising-like value $\nu\approx0.63$ in the limit $J_2/J_1\to\infty$. For $J_2/J_1\to 1/2$, critical exponents tends to values specific to a three-dimensional first-order transition in the finite-size scaling theory $\eta=-1$ (fig. \ref{fig4}) and $\nu=1/3$ (fig. \ref{fig5}).
\end{enumerate}

For the RG analysis, one should obtain the corresponding Ginzburg -- Landau functional from the lattice model (\ref{model}). To make this, we perform the small momentum expansion near the both minima $\mathbf{q}=(\pi,0,\pi)$ and $\mathbf{q}=(0,\pi,\pi)$, and replace the constrain $|s|=1$ by the additional potential $U(s)=m s^2+\lambda s^4$. Introducing the fields
\begin{equation}
    \phi=s|_{\mathbf{q}\approx(\pi,0,\pi)}+s|_{\mathbf{q}\approx(0,\pi,\pi)},\atop
    \psi=s|_{\mathbf{q}\approx(\pi,0,\pi)}-s|_{\mathbf{q}\approx(0,\pi,\pi)},
\end{equation}
we find after the field rescaling
$$
    F=\int d^dx \left(\left(\partial_\mu\phi\right)^2+\left(\partial_\mu\psi\right)^2+r(\phi^2+\psi^2)+\right.
$$
\begin{equation}
    \left.+u(\phi^4+\psi^4)+2v\phi^2\psi^2\right).
    \label{GLW}
\end{equation}
The region of the potential stability is
\begin{equation}
    u>0,\quad v>-u.
\end{equation}
In the 1-loop approximation in the minimal substraction scheme, the RG equations read
\begin{equation}
\partial_t u = -\varepsilon u + (9 u^2 + v^2)/2,\atop
\partial_t v = -\varepsilon v + (6 u v + 4 v^2)/2,
\end{equation}
where $t=-\ln(\mu/\Lambda)$, $\mu$ is the floating scale, and $\Lambda$ is the inverse lattice constant (UV cutoff). In fact, the RG equations are known at least in the 5-loop approximation \cite{Pelissetto03}, but taking into account the higher orders in $\varepsilon$ preserves the quantitative picture of the same.

\begin{figure}[t]
            \center
            \includegraphics[scale=0.70]{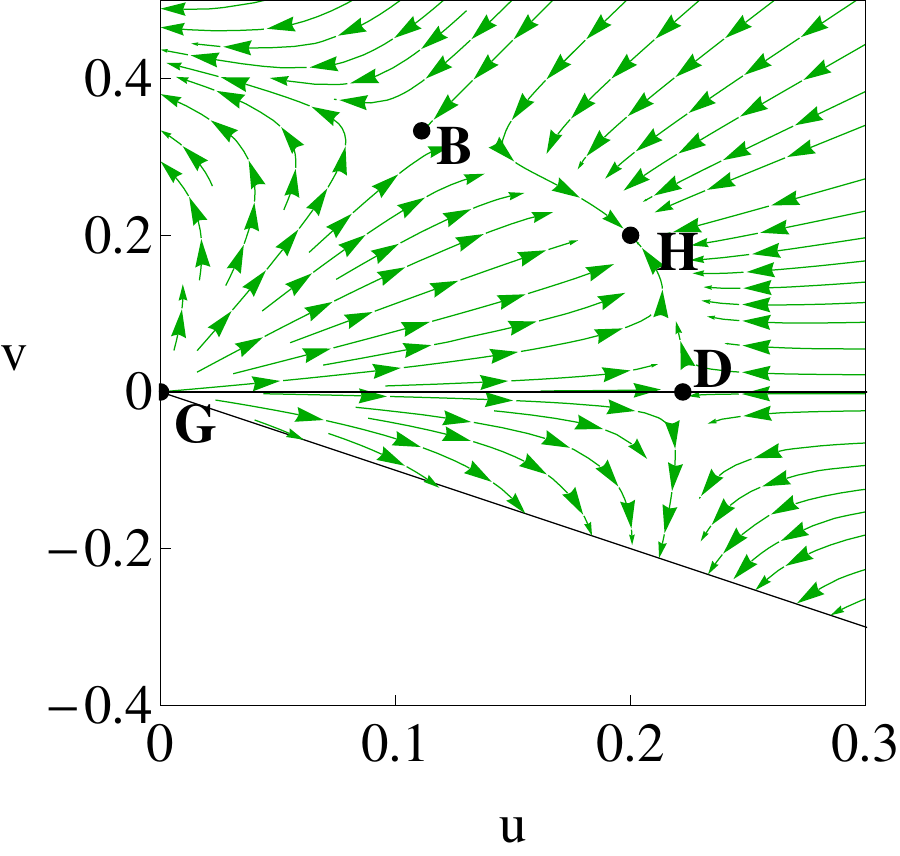}
\caption{\label{fig6}RG diagram.}
\end{figure}%
There are the four fixed point in the RG diagram (see fig. \ref{fig6}). Besides the trivial Gaussian fixed point $G$, one has the stable Heisenberg fixed point describing a phase transition from the universality class of the $O(2)$ model (see the functional (\ref{GLW}) with $u=v$), the unstable decoupled fixed point $D$ corresponding to two non-interacting Ising models, and the unstable so-called biconical fixed point $B$.

The coordinates of the initial point for the RG equations $\mu=\Lambda$ corresponding to the lattice model (\ref{model})  continuously depend on the exchange constants ratio $J_2/J_1$. In the limit $J_2/J_1\to\infty$, a RG trajectory starts from the vicinity of the decoupled fixed point. Since for all $J_2/J_1>1/2$ the critical behavior is not from the $O(2)$ class, we conclude that initial points locate in the sector $v<0$. All trajectories starting from this sector leave the stability region that means a first-order transition.

In addition, we have two more arguments for a first-order transition.
\begin{enumerate}
    \item[5.] In two dimensions, the transition line of the $J_1$-$J_2$ models is mapped to the corresponding line of the Ashkin-Teller model \cite{Jin12}, but in three dimensions, the Ashkin-Teller line is of the first order \cite{Arnold97}.

    \item[6.] Although the groups $\mathbb{Z}_2\otimes\mathbb{Z}_2$ and $\mathbb{Z}_4$ are not isomorphic, the true symmetry group breaking in the Ashkin-Teller model as well as in the $J_1$-$J_2$ model is $\mathbb{Z}_4$. The all four configurations shown in fig. \ref{fig1} can be sequentially obtained from one another by a lattice rotation in $\pi/2$. We know at least two models belonging to the same symmetry class: the 4-state Potts model which has a distinct first-order transition in three dimensions, and the 4-state clock (or planar Potts) model the critical behavior of which is described by the decoupled fixed point \cite{Hove03} (the Ising-like behavior).
\end{enumerate}

In conclusion, we note that our main statement on a first-order transition for all $J_2/J_1>1/2$ still relates to the special three-dimensional case $J/J_1=1$. When $0\leq J/J_1\leq1$, the situation interpolates between two- and three-dimensional cases. We expect that at $J/J_1\lesssim0.1$, a transition for large $J_2/J_1$ becomes continuous. The exact position of the tricritical point is need to be estimated by additional investigations.
\medskip

This work was supported by the Theoretical Physics and Mathematics Advancement Foundation 'BASIS' (project No. 19-1-3-38-1).

\end{document}